# He-Mg compounds and helium-driven nonmetal transition in metallic magnesium

Yu S. Huang,[1] Hong X. Song,[1] Qi D. Hao,[1,2] Xiao L. Pan,[1,2] Dan Wang,[1,2] Hao Wang,[1] Y. F. Wang,[1] Y. Sun,[1,†] Hua Y. Geng[1,3,*]

[1] *National Key Laboratory of Shock Wave and Detonation Physics, Institute of Fluid Physics, China Academy of Engineering Physics, Mianyang, Sichuan 621900, P. R. China;*

[2] *Institute of Atomic and Molecular Physics, College of Physics, Sichuan University, Chengdu 610065, P. R. China;*

[3] *HEDPS, Center for Applied Physics and Technology, and College of Engineering, Peking University, Beijing 100871, P. R. China.*

[†] To whom correspondence should be addressed. E-mail: ssunyyi00@163.com
[*] To whom correspondence should be addressed. E-mail: s102genghy@caep.cn

**ABSTRACT:** The polymorphism and mechanism of helium compounds is crucial for understanding the physical and chemical nature of He-bearing materials under pressures. Here, we predict two new types of He-bearing compounds, MgHe and $Mg_nHe$ (n = 6, 8, 10, 15, 18), being formed above 750 GPa by unbiased *ab initio* structure search. An unexpected bandgap is opened up in MgHe at as low as around 200 GPa. This is the first case of noble gas driven metal-nonmetal transition in all elements. The same mechanism is demonstrated also being applicable to other metallic elements, and making beryllium transform into a non-metallic state, a triumph that is impossible otherwise. Furthermore, the stability of the simple cubic phase of Mg (Mg-*sc*) is greatly enhanced by mixing with He, which lowers the critical pressure of pure Mg-*sc* from about 1.1 TPa down to 750 GPa to form ordered substitutional alloying phase of $Mg_nHe$ on a simple cubic lattice of Mg. This is the first report on Mg-based noble gas substitutional alloy, in sharp contrast to the conventional wisdom that He preferring interstitial sites. The observed striking influences of He demonstrate the rich physics and chemistry of He-bearing compounds under ultra-high pressures.

## I. Introduction

Helium (He) is one of the most abundant elements in the universe (up to 25% mass fraction), making it one of the primary constituents in many celestial bodies.[1,2] Therefore, the physicochemical properties of He under such extreme conditions is crucial for understanding their evolution, behavior, and lifespan. For example, the mixing and demixing of liquid He with hydrogen in the middle layer of Jupiter plays an important role in the three-layer model of that planet, which might have a surprising phenomenon called "helium rain".[3-5] It was reported that the abnormal luminosity of Saturn is also related to the properties of helium at high pressure.[6]

Helium atom has an ionization energy of 24.59 eV and a zero electron affinity[7,8], which hinders its chemical reaction with other elements. So far, it is known there are only four mechanisms for He to bind with other elements or materials, and none of them involve apparent chemical reactions. The first one is via pure Van der Waals (vdW) interaction. He can form weak vdW molecules or adsorbed to nanostructures at ultra-low temperatures.[9,10] Secondly, if the host object is positively or negatively charged, the electrostatic interaction induced polarization also helps He to form charged molecules such as $HeH^+$ and $FHeO^-$ at low temperature.[11,12] In the third mechanism, the volume reduction of the contribution to the enthalpy is the main driven force, such as those with nitrogen[13,14], other noble gases (NG) elements[15-17], and ammonia[18]. In the fourth mechanism, where He mainly acts as an inert space-occupier and must be inserted into the place between cations or anions to decreases the repulsive interactions.[19] This mechanism has been suggested in compounds of He formed with water[20-22], metals[23], metal oxides[24], and sulfides[25].

On the other hand, it was demonstrated both theoretically and experimentally that He and sodium (Na) can form a $Na_2He$ compound at pressures exceeding 113 GPa.[26] Noticeable change is caused by He insertion, which makes $Na_2He$ seem being different from all previously known inclusion compounds. It is noted that Na undergoes a metal-nonmetal transition at high pressure (i.e. TI19 → hP4 at around 273 GPa).[27] The helium enhanced interstitial quasi-atom (ISQ)[28] localization help to make this metal-nonmetal transition pressure down to 160 GPa. Recently, Liu *et al.* demonstrated that $Na_2He$ should be considered as the insertion of He into ionic compound with unequal number of cations and anions.[19] This picture becomes evident when writing the formula explicitly as $Na_2EHe$, where E stands for the electrons in ISQ. In this way, $Na_2He$ belongs to the fourth mechanism as mentioned above. The only difference is the presence of ISQ as anions in this electride makes the change by He insertion more prominent than in other He-bearing compounds of the same category. Therefore, an intriguing question is, whether is there another mechanism for He-bearing compounds beyond these four types or not.

Furthermore, it is well known that both lithium (Li) and Na are electride and transform into non-metallic phases at high pressure. In contrast, beryllium (Be), magnesium, and aluminum are also electride with pseudogap at high pressure, but their bandgap never can be opened up (in which only Mg becomes a semimetal when beyond 2.7 TPa). To the best of our knowledge, no NG-induced metal-nonmetal transition has been reported in all of these metals. One may wonder whether it is possible or not. To address these issues, we performed an unbiased evolutionary crystal structure prediction for the He-Mg system up to 1 TPa. A series of novel He-Mg compounds are discovered. The observed prominent charge transfer and hybridization between He and Mg atoms indicate He has partially lost its inertness and chemical reactions are indeed involved, which provides the fifth mechanism for He-bearing compounds under high pressure.

## II. Computational Method

Crystal structure prediction (CSP) for the He-Mg system is carried out by using the *ab initio* evolutionary algorithm as implemented in USPEX[29,30] from 700 to 1000 GPa with a pressure interval of 50 GPa. The enthalpy of formation is calculated to determine whether or not the structure is thermodynamically stable. An initial population size of 60 created randomly is employed for the first generation, and all succeeding generations have a population size half of that. Furthermore, 40% structures of each generation with the worst enthalpy are discarded, and new structures are added. The whole structure search stops after 40 generations if all stable structures have already been found earlier than the last generation.

All structure optimization and electronic structure calculations are performed by using density functional theory (DFT)[31,32] as implemented in Vienna *ab initio* Simulation Package (VASP)[33]. The Perdew-Burke-Ernzerhof (PBE)[34] parameterized exchange-correlation functional within the generalized gradient approximation (GGA) and the projector augmented wave (PAW)[35] method are used to describe the ion-electron interaction. The valence configuration for Mg and He are $2s^22p^63s^2$ and $1s^2$, respectively. As shown in Fig. S8 in SM[36], a careful test of core overlap is performed to ensure the validity of the PAW pseudopotential of Mg and He under such extreme conditions. To converge the total energy better than 1 meV/atom, a kinetic energy cut-off of 1000 eV for the plane-wave basis and a Γ-centered reciprocal space mesh with a resolution of 2π×0.016 Å$^{-1}$ for the k-points sampling are employed. The dynamical stability of all structures are examined by phonon calculations that using PHONOPY[37] and our homemade MyPHON[38] code. The maximally-localized Wannier functions are calculated by Wannier90.[39]

## III. Results and discussion

### A. Novel structures in He-Mg system

In He-Mg system, we find six stable structures spanning a wide range of stoichiometry by using CSP, in contrast to previous thought that no stable He-Mg compounds at high pressure. The calculated formation enthalpy of these structures are shown as the convex hull in Fig. 1(a). More detailed CSP results can be found in Fig. S1 of SM. With increasing pressure, $Mg_6He$ is the first compound that becomes stable at 750 GPa, followed by other compounds subsequently. The structure phase diagram is shown in Fig. 1(b), which provides the details of their space group, and suggests that there is no phase transition within their respective stable pressure ranges. Their dynamical stability is further confirmed by the calculated phonon spectra that have no imaginary frequency (see Fig. S2 of SM).

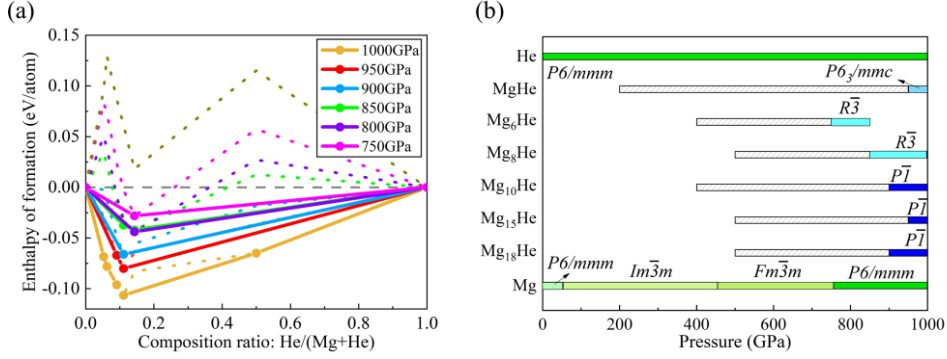

**Figure 1.** The CSP results of He-Mg system within the pressure range between 700 and 1000 GPa. (a) Enthalpy of formation for the six predicted stable structures. (b) Structural phase diagram of the six compounds and the elemental He and Mg, where the hatched region indicates the metastable pressure range.

Based on the features of structural geometry, these six new structures can be classified into two groups: MgHe and the other Mg-rich compounds $Mg_nHe$. The former has an anti-NiAs structure with the space group *P6$_3$/mmc*, which can be viewed as a simple hexagonal (*sh*) sublattice of He inserted into a hexagonal closed packed (*hcp*) sublattice of Mg, as depicted in Fig. 2(a). In contrast, the structures in the latter group can be considered as Mg-based substitutional alloys, in which the matrix is a simple cubic (*sc*) lattice formed by Mg atoms, with some of them are substituted by He atoms. For example, the structure of $Mg_6He$ is a supercell of *sc* matrix of Mg, and the substituted He atoms are on the hexagonal dense packing planes, with its normal vector oriented along the [111] direction. These planes adopt an ABCABC… type of stacking. The schematic diagram is illustrated in Fig. 2(b). It should be noted that because of its chemical inertness and small atomic size, helium is usually preferred the interstitial sites when entering the host matrix.[40] To the best of our knowledge, this is the first reported Mg-based substitutional alloy with NG atoms. It is totally against the conventional wisdom. More details about the substitution and the structure parameters can be found in Fig. S3 and Table S1 of SM.

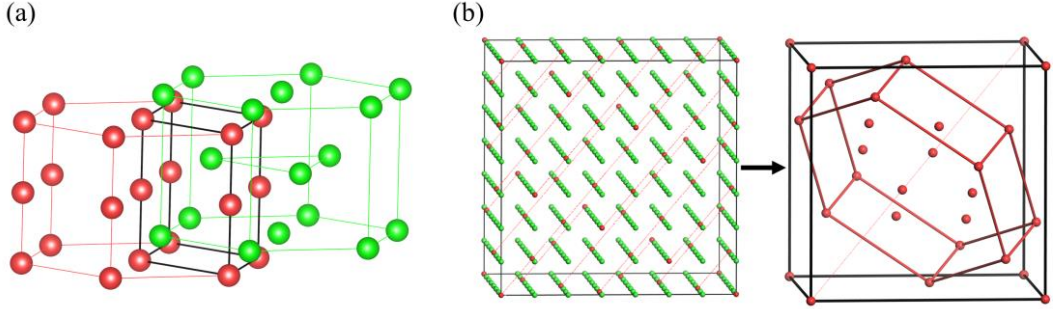

**Figure 2.** The structure of (a) MgHe and (b) Mg$_6$He, with the red and green spheres representing the He and Mg atoms, respectively. The red, green, and black line in (a) shows the He sublattice, Mg sublattice, and MgHe primitive cell, respectively. The red dashed lines in (b) indicate the [111] direction of the Mg$_6$He substitutional alloy.

**B. Origin of stability**

Considering Mg$_8$He has the deepest formation enthalpy ($\Delta H_f \approx -0.106$ eV/atom at 1 TPa) and MgHe is an insulating phase, we thus focus on MgHe and Mg$_8$He to discuss their mechanism of stability. Figure 3(a)-(b) plots the individual contributions to $\Delta H_f$ as a function of pressure. The sharp variation in $\Delta E$ and $P\Delta V$ within the pressure range of 700 to 750 GPa as shown in Fig. 3(a)-(b) is due to the *fcc-sh* transition in the end reference material of Mg element. Obviously, the main contribution to the stability is from the reduction of the volume (manifests as negative $P\Delta V$ term in $\Delta H_f$) in the studied pressure range. Actually, the reduction of volume here is not contributed by the compression of the vdW space of He only. Otherwise the He-Mg compounds should appear at a much lower pressure, like other He-bearing compounds (usually at a scale of ~ 100 GPa). The large volume reduction is in fact coming from the ISQ formed in the interstitial sites in MgHe and Mg$_8$He. As shown in Fig. S6 of SM, the calculated electron localization function (ELF)[41] indicates that MgHe and Mg$_8$He belong to high-pressure electrides (HPEs)[42,43]. The highly localized ISQ formed there significantly reduces the charge density in other places (shown in Fig. 3(c)-(d)). For example, in Mg sublattice of MgHe, the ISQ formed at the center of triangular dipyramid (denoted as ISQII) significantly reduce the charge density at the cell edges (denoted as ISQI, where He atoms will be inserted into), which lowers the repulsion with the inserted He, and resulting in a large volume reduction by comparison to pure Mg and He.

Furthermore, chemical interaction with He also comes into play, which relatively lowers the internal energy. For example, an amount of charge about 0.28e is accumulated on He in MgHe (about 0.38e on He in Mg$_8$He), which effectively reduces the ion-ion repulsion if compared to the rigid MgEHe model of the fourth mechanism (here the letter "E" refers the electrons in the highly localized in ISQII site). The actual chemical formula of MgHe is Mg$^{2+}$(E$^{1.5-}$He$^{0.5-}$). The excessive charge on He atoms lowers down the Madelung energy with respect to the rigid MgE model. More importantly, insertion of He increases the nominal valence state of both Mg (from 1+ to 2+) and ISQII (from 1- to 1.5-), which enhances the polarity of this ionic compound, and decreases the electrostatic interaction energy further. Finally, there is strong electronic hybridization between He and Mg in MgHe, which also lowers down the band energy (see Fig. 4(a)). Generally speaking, it is the combination of these three effects, i.e. (i) the localization of electrons to ISQ to leave more space for He, (ii) excessive charge accumulation on He atoms, and (iii) the hybridization between He and Mg, that stabilizes MgHe. The same is true for $Mg_8He$ and other Mg-rich helium compounds.

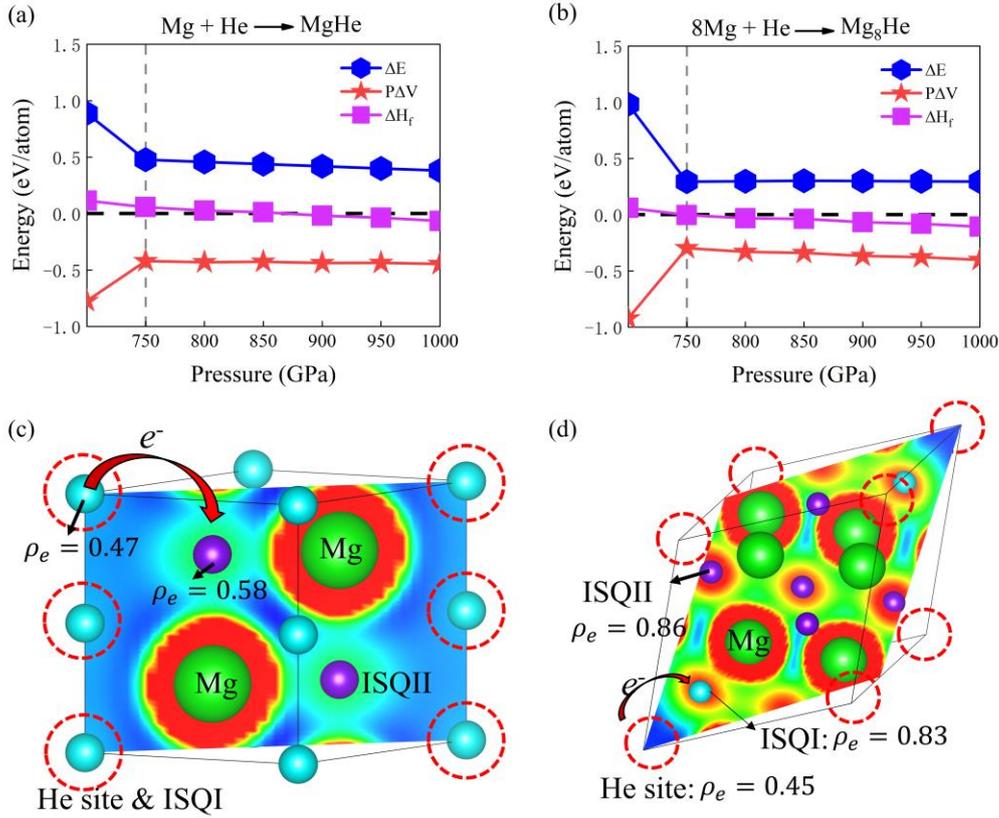

**Figure 3.** (a) The enthalpy of formation ($\Delta H_f$), the contribution from the change of energy ($\Delta E$), and of $P\Delta V$ for MgHe as a function of pressure. (b) $\Delta H_f$, $\Delta E$, $P\Delta V$ for $Mg_8He$. (c) and (d) is the charge density of Mg sublattice of MgHe and of $Mg_8He$, respectively. The large green balls respect the Mg atoms, and the small blue and purple balls represent the ISQI and ISQII, respectively. The red dashed circles represent the sites where He will be inserted into. For MgHe, ISQII attracts electrons away from ISQI (He site). In $Mg_8He$, both of the ISQI and ISQII attract electrons away from He sites.

**C. Chemical bonding nature**

To understand the chemical nature and the influence of the inserted He atoms, we calculated the projected electronic density of states (PDOS) for MgHe and Mg$_8$He at 1 TPa, respectively. For comparison, the PDOS of the pure Mg with *sc* structure (Mg-*sc*) and *hcp* structure (Mg-*hcp*) are also presented.

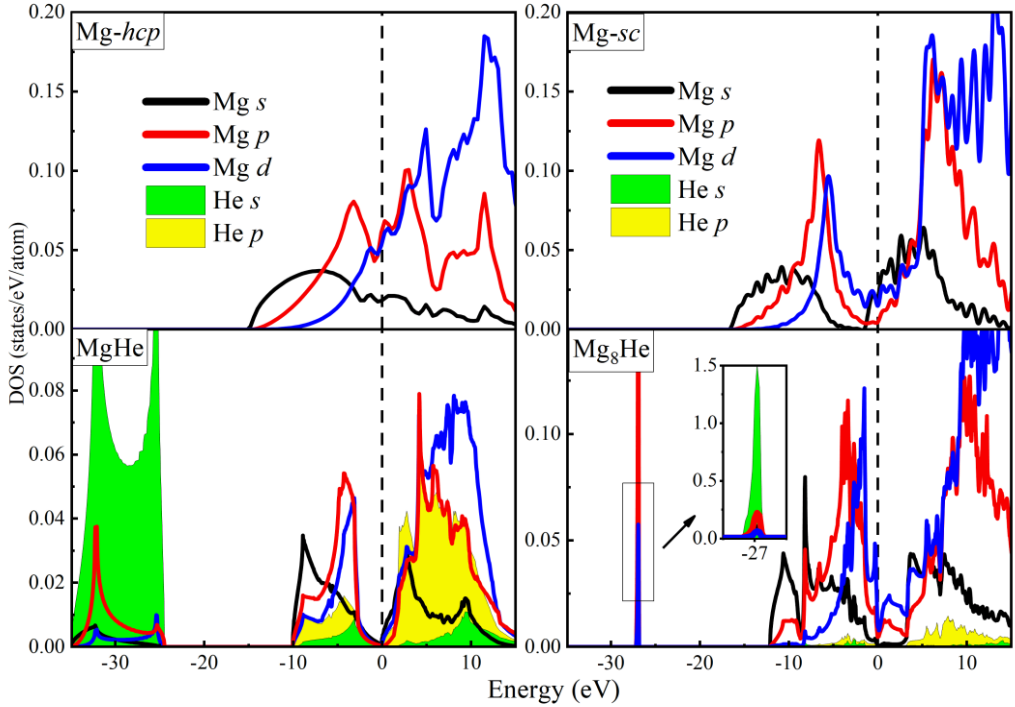

**Figure 4.** The calculated PDOS for Mg-*hcp*, MgHe, Mg-*sc* and Mg$_8$He at 1 TPa with PBE level. The structure of Mg-*hcp* and Mg-*sc* is the sublattice of MgHe and Mg$_8$He, respectively.

As shown in Fig. 4, the *p* and *d* states of Mg in the Mg$_8$He cross the Fermi level ($E_F$), suggesting its metallic character, while the PDOS at $E_F$ for MgHe is very small, showing an almost opening bandgap at 1 TPa. Intriguingly, we found an unexpected bandgap of MgHe at lower pressures (in the range from 200 to 800 GPa). The maximum bandgap is 0.59 eV at 400 GPa. Since the PBE functional always underestimates the bandgap,[44] we performed band structure calculation again by using more accurate GW method, and confirmed the nonmetallic nature of MgHe in the whole stable pressure range (as shown in Fig. S5, the bandgap is 2.06 eV at 1 TPa). At higher pressure, GW method shows that MgHe maintains its nonmetallic nature at least up to 3 TPa. This is the first report for NG atoms driven metal-nonmetal transition in Mg.

It is evident from Fig. 4 that the influence of He insertion on the electronic structure of MgHe is significant. By comparing to the pure Mg-*hcp* structure, there is strong hybridization between the 3*s*, 3*p*, and 3*d* states of Mg and the 2*s* and 2*p* states of He, and the bandgap is opened between the bonding and anti-bonding states of them at the Fermi level. Interestingly, the 3*p* orbitals of Mg could even extend to the location of He, and hybridize with the 1*s* orbitals of He (a semi-core state). Such hybridization might explain why He atoms in MgHe shows an abnormal broadened 1*s* state spanning from -35 to -24 eV (see Fig. 4). By contrast, the 1*s* state of He in other compounds is highly localized. To confirm the chemical nature of this phenomenon, the absolute square of the electron wave-function of Gamma point in MgHe was calculated in the energy range of -35 to -24 eV (see Fig. 5(a)). The wave-functions of MgHe indicate that two 1*s* orbitals of adjacent He atoms hybridize into a pair of bonding and antibonding states, which have broadened the width of the He-1*s* state from 0.3 eV (in $Mg_8He$) to about 11 eV (in MgHe). Meanwhile, the wave-function of MgHe shows a large probability density around Mg atoms. However, the wave-function does not provide a clear proof of the bonding between Mg and He. Therefore, we further caclulated the maximally-localized Wannier function (MLWF) of MgHe to confirm the bonding. As shown in Fig. 5(b), the sharing of a common MLWF between Mg and He atoms clearly demonstrates the bonding between Mg and He atoms. The participation of the He-$1s$ electrons in hybridization indicates that helium has lost its inertness, which further confirms the chemical reaction in MgHe.

For $Mg_8He$, the PDOS (in Fig. 4) exhibits negligible He states near $E_F$, suggesting the interaction between He and Mg is weaker in this case. Nevertheless, the He-$2p$ state presents and has a weak hybridization with Mg. The semi-core He-$1s$ state also weakly hybridizes with the Mg $3p$ states. These effects narrow the valence band from -18 eV in pure Mg-$sc$ phase to -12 eV in $Mg_8He$. In contrast to MgHe, the He atoms in $Mg_8He$ display an isolated atom-like $1s$ states at -27 eV. This difference is easy to understand, since as the concentration of Mg in $Mg_nHe$ increases, the distance between He atoms increases and prevents the $s$ orbitals of He atoms from overlapping or hybridization. However, it should be noted that this little amount of He could significantly lower the stabilization pressure of simple cubic phase of Mg from 1.1 TPa to 750 GPa, a remarkable reduction of 32%.

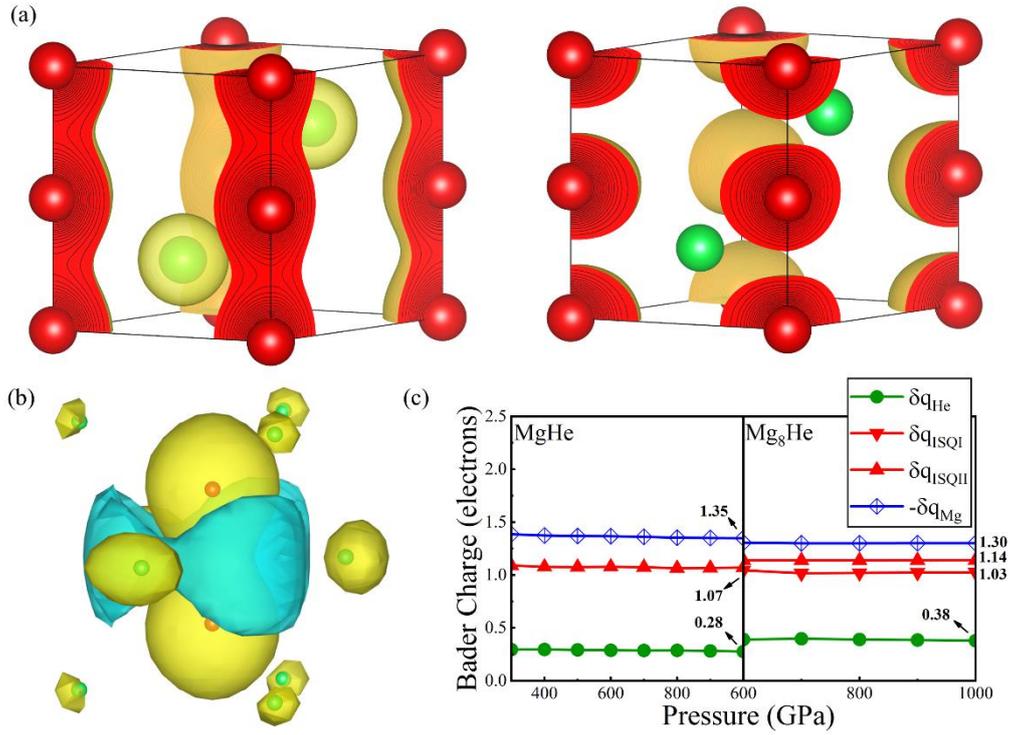

**Figure 5.** (a) The absolute square of wave-function for MgHe in the range of -35 eV to -24 eV. The red and green balls represent the He and Mg atoms, respectively. (b) The MLWF for MgHe, red and green balls represent He and Mg centers, respectively. (c) Bader charge of MgHe and $Mg_8He$ as function of pressure, in a unit of e.

The Bader's atom-in-molecule charge analysis[45] provides further insight for the bonding nature in these systems. For MgHe, the main charge transfer is from Mg to ISQ, which counts about 1.07e per site at 1 TPa (see Fig. 5(c)). Each He atom captures 0.28e, which is relatively a large gain if compared to other He compounds, e.g. 0.15e in $Na_2He$ [26] and 0.03 in $He@H_2O$[20]. Such large charge accumulation onto He demonstrates that there is indeed a chemical interaction between He and the Mg matrix, which makes MgHe behavior more like a magnesium salt. Indeed, we found more charges are pulled away from Mg atoms after inserting He, e.g. $q_{\text{Mg in Mg}-hcp} -$ $q_{\text{Mg in MgHe}} = 0.19\text{e}$ at 1 TPa. This might also be one of the reasons that enhance the nonmetallic feature of MgHe.

In contrast, the situation is a little bit different for the $Mg_8He$. There are two positions for electrons to localize. The ISQI formed near the He substitution site, and the ISQII formed in the center of Mg cube, far away from the He site (see Fig. S6). The Bader charges of ISQI and ISQII are 1.03e and 1.14e at 1 TPa, respectively. The accumulated charge of ISQII is smaller than the charge of ISQ in Mg-*sc* (about 1.33e), indicating that He attracts electrons from the ISQ site even when it is far away from it. In fact, the Bader charge of ISQII shows a slight decrease, when the He atom is inserted into the Mg sublattice of $Mg_8He$ (from 1.18e to 1.14e). More details about the charge transfer can be found in Fig. S6 of SM. In a similar way to MgHe, He atoms also capture surrounding electrons (with about 0.38e at 1 TPa); this phenomenon indicates that the substitution of Mg by He can not be viewed as a point defect in ionic compounds, where a positive charge is expected, like Xe impurity in $UO_2$.[46,47]

**D. Substitutional rules in $Mg_nHe$**

For $Mg_nHe$ (n=6, 8, 10, 15, 18) substitutional alloys, we found they obey a general substitutional rule. For example, both $Mg_6He$ and $Mg_8He$ have a similar He sublattice with a space group $R\bar{3}m$. Hence the substitution in $Mg_6He$ and $Mg_8He$ can be described using the same super-cell of Mg-*sc* which is then substituted by He sublattice that is identical in space group but with different size. For $Mg_6He$ and $Mg_8He$, this rule can be characterized by the number of Mg-*sc* units contained along the [111] direction in the He sublattice (see Fig. S4 (a)). In this way, $Mg_6He$ and $Mg_8He$ contain 1 and 3 Mg-*sc* units, respectively. This observation allows us to construct structures with higher composition of Mg, e.g. Mg$_{19}$He and Mg$_{27}$He that containing 5 and 7 Mg-*sc* units, respectively. Indeed, their formation enthalpy is very close to the convex hull, indicating both of them might be thermodynamically stable (Fig. S4 (b)).

For Mg$_{10}$He, Mg$_{15}$He and Mg$_{18}$He, the same substitutional rule might apply. The difference is that in this case the space group of the He sublattice is $P\bar{1}$ instead of $R\bar{3}m$. The existence of a substitutional rule in Mg$_n$He suggests that there is long-ranged ordering in the He sublattice in these alloys, which is quite unique and totally unaware of before. This argument is also supported by following observation. We noticed that even though the $\Delta H_f$ of all of these alloys distributed on the convex hull along an almost straight line, they in fact are not disordered solid solutions but being ordered phases. For example, when the distribution of He atoms on the Mg-*sc* matrix becomes slightly random, the formation enthalpy will increase to at least approximately 0.15 eV/atom, suggesting a long-range interaction is also one of the driven forces to stabilize this system.

**E. Helium-driven insulating mechanism**

For the MgHe, the insulating mechanism can be understood from two aspects: (i) He atoms repel electrons into the ISQs, making the electrons more localized and valence band narrowed (see Fig. 4); (ii) He 1*s*, 2*s*, and 2*p* orbitals have strong hybridization with Mg, which helps to open the bonding-antibonding gap of E$_F$. Considering that other NG element have similar behavior as He, it is natural to assume that generally NG also can drive metal-nonmetal transition. To verify that, we replaced

He in MgHe with another NG atom neon (Ne). The calculated electronic structure of MgNe shows that, as we predicted, Ne indeed help to open the bandgap of Mg (the gap is about two folds bigger, with ~4 eV, see Fig. S7 in SM).

In order to illustrate its generality, we also consider another metal. It is well known that beryllium cannot be compressed into a nonmetallic phase at any pressure. One might wonder whether the same NG-driven metal-nonmetal transition can be induced in Be or not. To this end, we tried to insert Ne into Be in the same structure. The results show that a bandgap of ~0.1 eV (at the PBE level) is opened up in Be by Ne, which is similar to MgHe. This demonstrates that the above-mentioned helium atom driven metal-nonmetal transition is quite general, and can be applied to other NG atoms and metallic elements. This insight opens a new avenue to diverse the physicochemical properties of NG-bearing compounds.

Finally, we should emphasize that we have illustrated that He can react with Mg to form stable compounds in terapascal pressure range. Such pressures are commonly found in the interiors of many celestial bodies, such as the mantle-core boundary of Jupiter (approximately 4.2 TPa) and Saturn (around 1 TPa) where He and Mg are among the main constituent elements.[48,49] Therefore, in the interior of planets like Jupiter and Saturn, the rocky core may capture He from their atmosphere. In fact, the atmosphere of Saturn is indeed He-deficient;[6] and the above discovered reaction between He and Mg at relevant pressures might provide a possible explanation for this phenomenon. Additionally, the reaction between He and Mg also suggests that within the interior of these giant gas planets, the rocky core might undergo reactions with the He, leading to core erosion, and changing the composition distribution around it.

## IV. Conclusion

In summary, unbiased *ab initio* crystal structure prediction was performed for He-Mg system. It was demonstrated that He reacts with Mg to form unexpected insulating MgHe compound and substitutional $Mg_nHe$ alloys in the range of 750 GPa to 1 TPa. The mechanism of their structural stability is resultant from the charge localization to ISQ to leave space for He insertion, prominent charge transfer to He, and electronic hybridization between Mg and He orbitals. This work also revealed a novel chemical role played by He that never been imagined: it drives metallic Mg into an insulating phase. The same property also holds by Ne, which also drives both Mg and Be into a nonmetallic state, the first realization of such state in both metals. We also found that a little bit He does a lot to Mg, by lowering the stable pressure of the simple cubic phase of Mg from about 1.1 TPa[50] to 750 GPa, and forming unconventional alloys with helium. It is found the substitutional $Mg_nHe$ alloys obey a general construction rule, implying there is long-ranged interaction between the He atoms on the substituted sublattice. using this rule, two new substitutional compounds of $Mg_{19}He$ and $Mg_{27}He$ were successfully predicted. These findings have profound impact for understanding the interior of planets such as Jupiter and Saturn. They also exhibit the rich physics and chemistry of He at terapascal pressure range. It might stimulate experimental attempts into this territory, which just becoming possible with recently emerging ultra-high pressure techniques such as secondary micro-anvil of DAC.[51,52]

## Conflicts of interest

The authors declare no competing interests.

## Data availability

All data can be found in the manuscript and supporting materials.

## Code availability

The VASP code used to calculate electronic structure and structural optimization is a commercial software supported by VASP Software GmbH, while others are open-source codes which can be obtained from their respective official websites.

## Acknowledgments

This work was supported by the National Key R&D Program of China under Grant No. 2021YFB3802300, the National Natural Science Foundation of China under Grant No. 12372370 and the NSAF under Grant No. U1730248. Part of the computation was performed using the supercomputer at the Center for Computational Materials Science (CCMS) of the Institute for Materials Research (IMR) at Tohoku University, Japan.

## CRediT authorship contribution statement

Yu S. Huang: Investigation, Methodology, Writing - original draft, Writing - review & editing. Hong X. Song: Methodology, Writing - review & editing, Qi D. Hao: Methodology, Writing - review & editing, Xiao L. Pan: Methodology, Writing - review & editing, Dan Wang: Methodology, Writing - review & editing, Hao Wang: Methodology, Writing - review & editing. Y. F. Wang: Methodology, Writing - review & editing, Y. Sun: Methodology, Writing - review & editing, Hua Y. Geng: Idea conceiving, Project design, Writing, Reviewing, and Editing, Supervision, Project administration, Software.

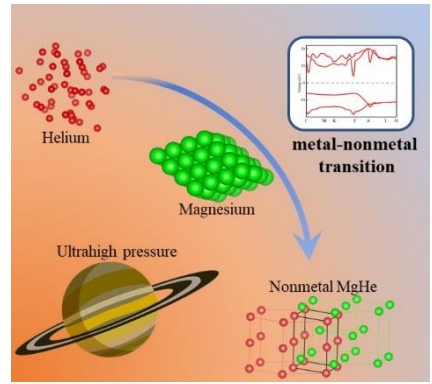

The noble gas (NG) helium (He) has been shown to be non-noble under high pressure. In extreme environments, such as giant planets, the He in the atmosphere can be captured by the rocky core and react with elements within it. Here, we find He reacts with magnesium (Mg), leading to the transform of metallic Mg into a nonmetallic phase. These phenomena suggest that the capture of He by the rocky core of a planet may alter its chemical properties, affecting its material distribution and evolution.

# Supporting materials for "He-Mg compounds and helium-driven nonmetal transition in metallic magnesium"

Yu S. Huang,[1] Hong X. Song,[1] Qi D. Hao,[1,2] Xiao L. Pan,[1,2] Dan Wang,[1,2] Hao Wang,[1] Y. F. Wang,[1] Y. Sun,[1†] Hua Y. Geng[1,3*]

[1] *National Key Laboratory of Shock Wave and Detonation Physics, Institute of Fluid Physics, China Academy of Engineering Physics, Mianyang, Sichuan 621900, P. R. China;*

[2] *Institute of Atomic and Molecular Physics, College of Physics, Sichuan University, Chengdu 610065, P. R. China;*

[3] *HEDPS, Center for Applied Physics and Technology, and College of Engineering, Peking University, Beijing 100871, P. R. China.*

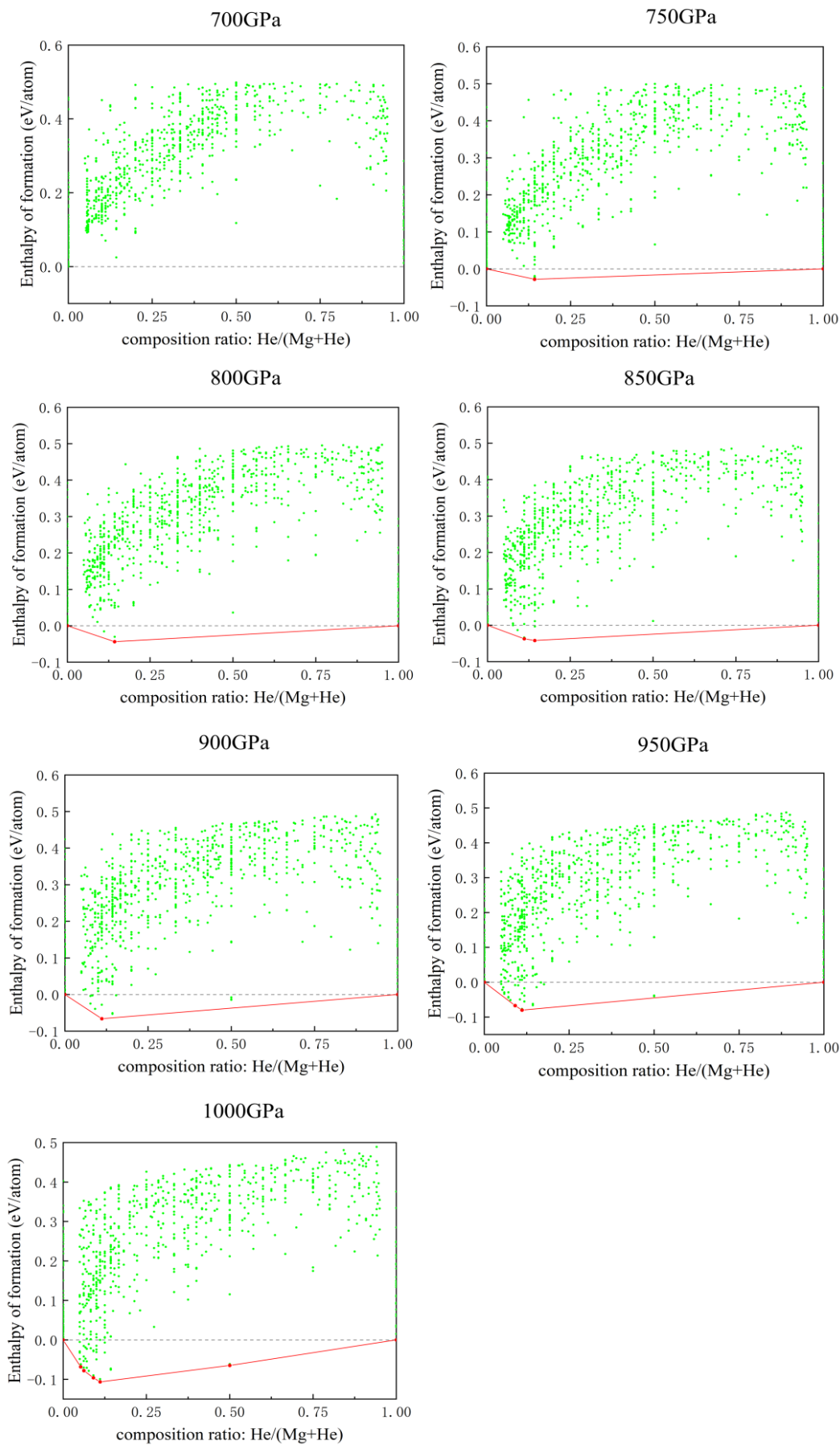

**Figure S1.** The result of CSP calculations. A preliminary test shows that several stable structures exist at pressures exceeding 700 GPa. Therefore, we conducted a series of thorough and precise CSP calculations within the pressure range from 700 to 1000 GPa. The results indicate that the system tends to form Mg-rich alloys, including both stable and metastable structures, except a special MgHe compound. Our CSP calculations employed unit cells of containing up to 20 atoms.

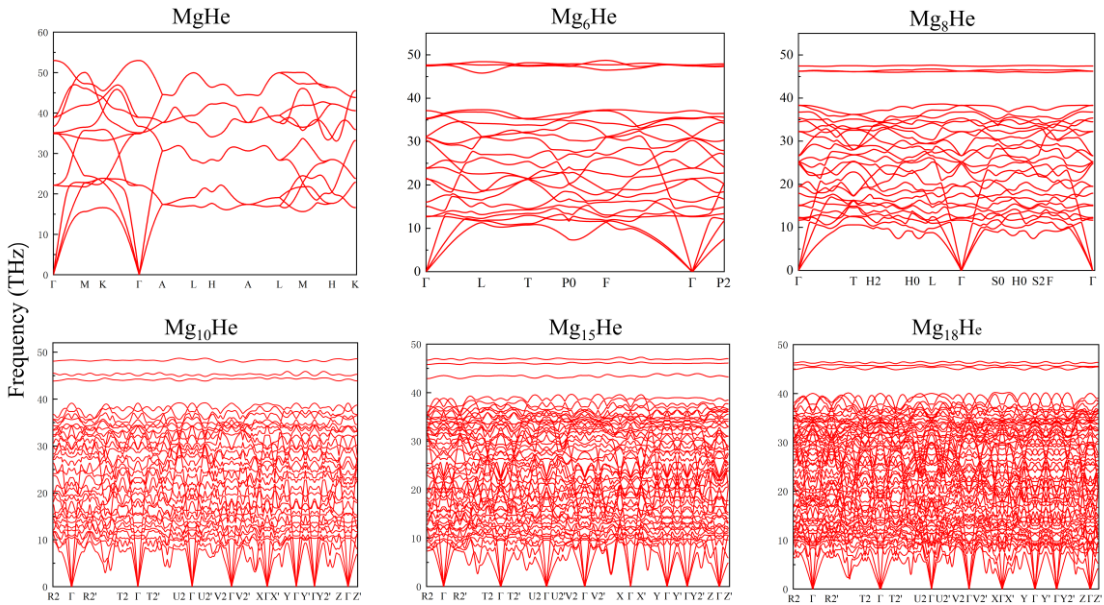

**Figure S2.** The phonon dispersion spectrum of MgHe, $Mg_6He$, $Mg_8He$, $Mg_{10}He$, $Mg_{15}He$, and $Mg_{18}He$ at 1000 GPa, respectively. The dynamical stability is confirmed by the phonon dispersion spectrum that without any imaginary frequency. Additionally, we extended the phonon calculations down to much lower pressures, and the results determine the dynamical stability range of these six structures. Their respective stable and metastable pressure intervals are presented in Fig. 1 of the main text. We also note the localized vibrational modes of He in Mg-rich phases, as well as the strong mixing

of He modes and Mg modes in MgHe.

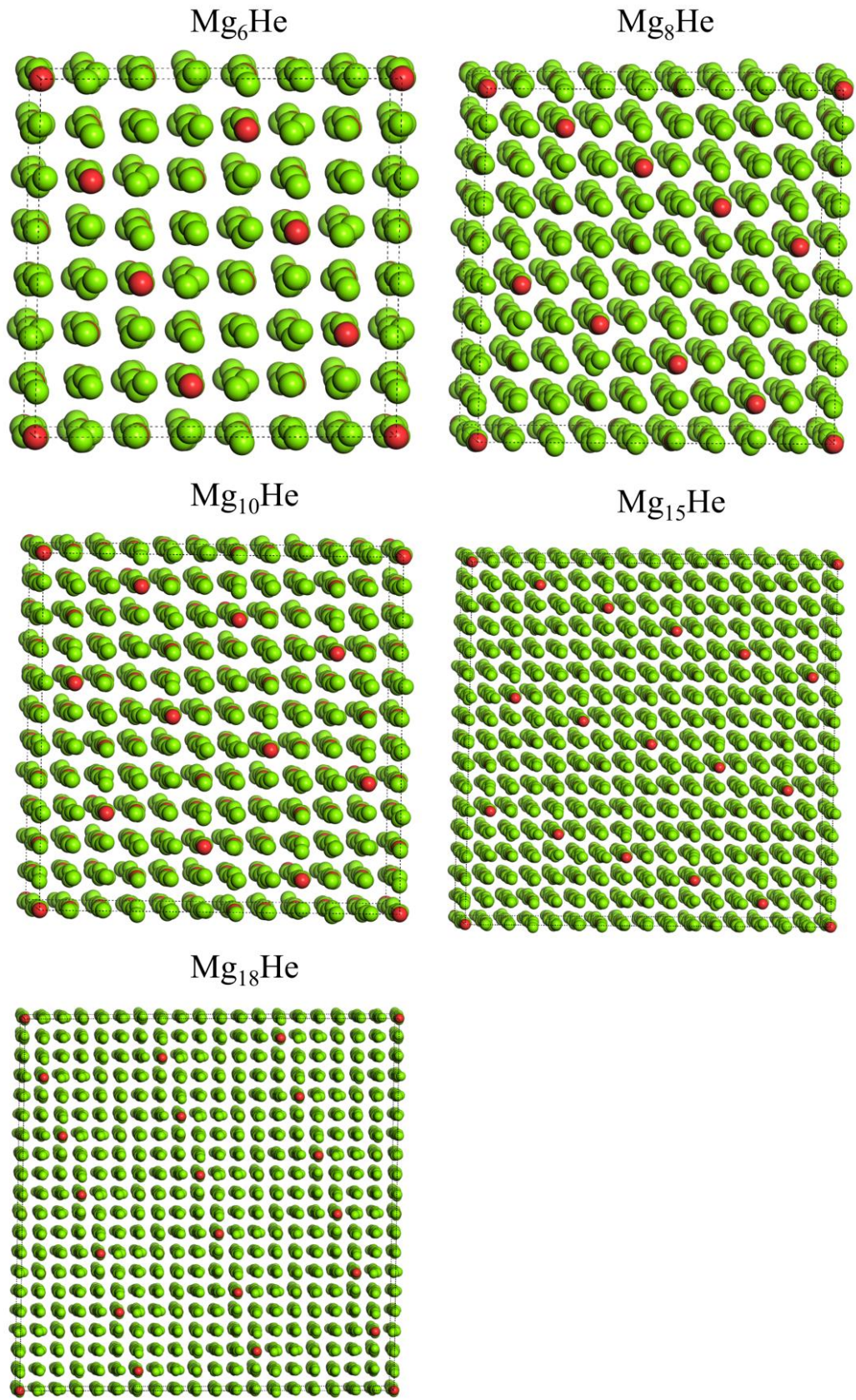

**Figure S3.** Structures of the stable Mg-based substitutional $Mg_nHe$ alloys. The red and green spheres represent He and Mg atoms, respectively. The structural distortion of the Mg-*sc* lattice induced by He substitution can be clearly seen. The smaller the *n* is, the less the distortion of the lattice becomes. In these $Mg_n$He, $Mg_6$He and $Mg_8$He share similar substitutional characteristics, which can be viewed as an *hcp* sublattice formed by He atoms replacing Mg atoms along the [111] direction. The only difference between $Mg_6$He and $Mg_8$He lies in a larger He sublattice in $Mg_8$He. For $Mg_{10}$He, $Mg_{15}$He, and $Mg_{18}$He, their structure can be approximately regarded as distorted *sh* sublattice formed by He atoms that replacing Mg atoms along different crystal directions.

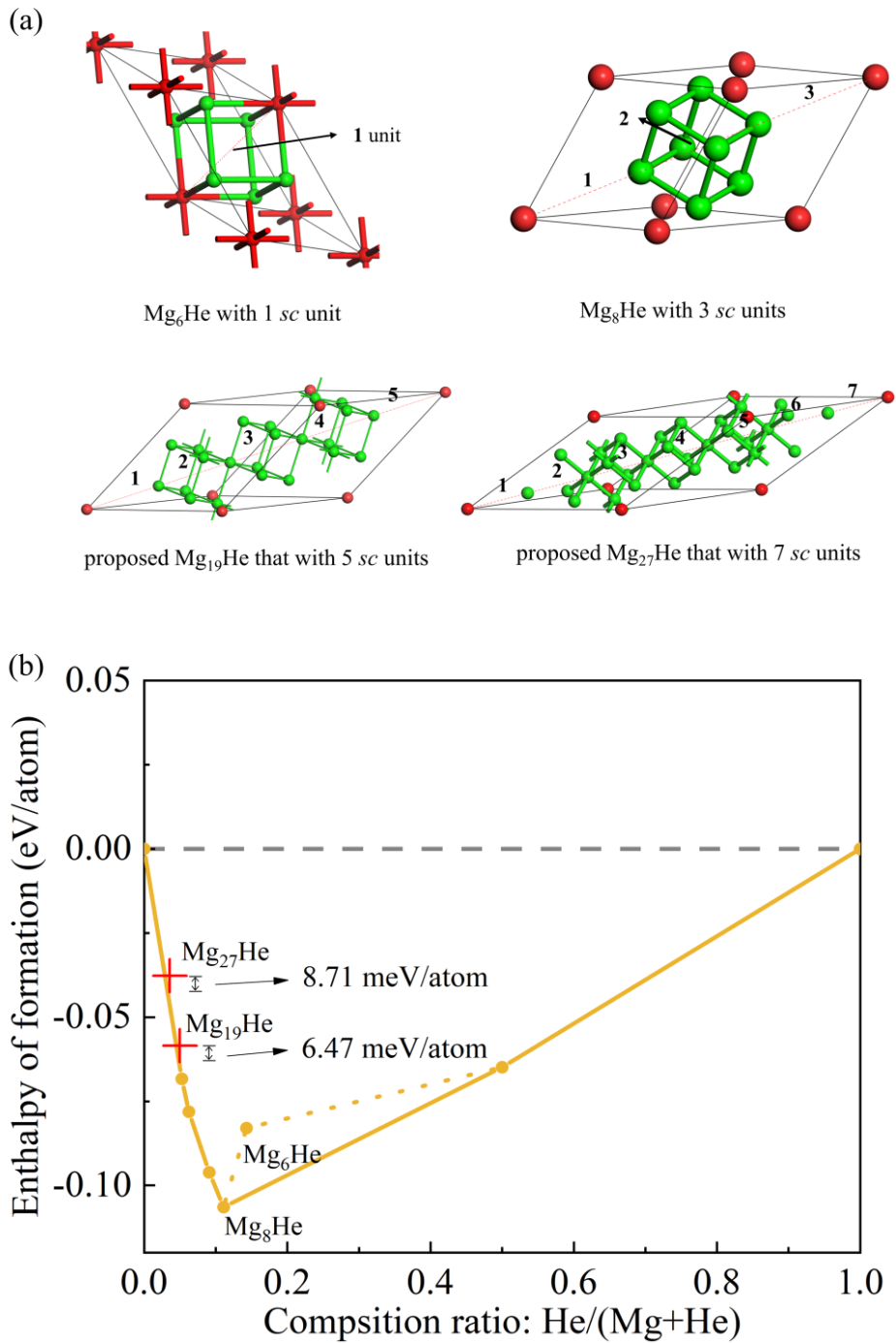

**Figure S4.** (a) The primitive cell of Mg$_6$He, Mg$_8$He, and the proposed Mg$_{19}$He and Mg$_{27}$He, respectively. The red and green balls represent He and Mg atoms, respectively. The red dashed line marks the [111] direction of the He sublattice. It is evident that Mg$_6$He and Mg$_8$He have 1 and 3 Mg-*sc* units along the [111] direction, respectively. Following this rule, Mg$_n$He alloys with greater n, i.e. Mg$_{19}$He and Mg$_{27}$He, are constructed by using 5 and 7 Mg-*sc* units along the [111] direction, respectively. (b) The formation enthalpy of Mg$_6$He, Mg$_8$He, Mg$_{19}$He and Mg$_{27}$He at 1 TPa. Especially, the formation enthalpy of the proposed Mg$_{19}$He and Mg$_{27}$He is -0.0585 and -0.0376 eV/atom, respectively, which are very close to the convex hull, with a small distance of only 6.47 meV/atom and 8.71 meV/atom, respectively. They might also be thermodynamically stable. Their dynamical stability is confirmed by *ab initio* molecular dynamics simulation performed at 300 K and higher temperatures.

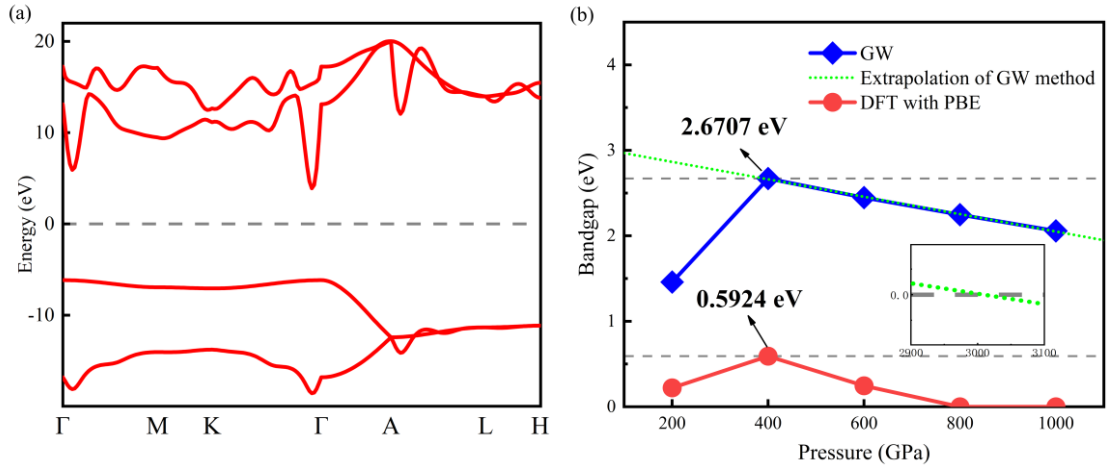

**Figure S5.** (a) The band structure of MgHe calculated by GW method at 1 TPa, which exhibits a wide bandgap of approximately 2.06 eV. The flat valence bands reveal the localized nature of valence electrons. (b) The bandgap of MgHe as a function of pressure calculated by GW and DFT with PBE functional, respectively. The bandgap maximum evaluated by both the PBE and GW methods occurs at 400 GPa. The PBE predicts bandgap opening below 800 GPa, while the GW results suggest that the bandgap holds up to at least 3 TPa by extrapolation.

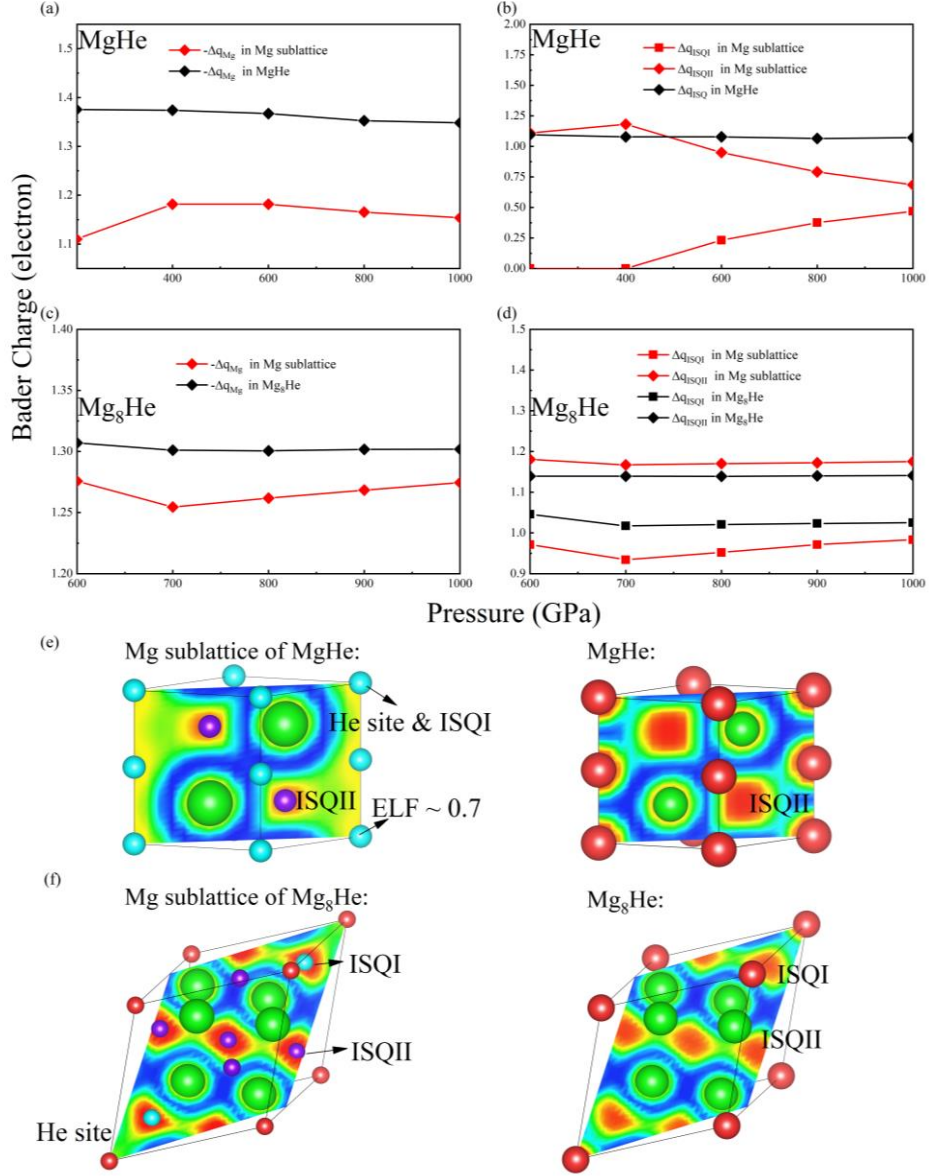

**Figure S6.** (a) Variation of the Bader charge on Mg atoms in MgHe and its Mg sublattice, respectively. It can be seen that Mg loses about 0.2e electrons after He insertion. (b) Variation of the Bader charge on the ISQ at the triangular dipyramid centers in MgHe and its Mg sublattice, respectively. It exhibits that He pushes electrons from ISQI to ISQII sites. (c) Bader charge on Mg in $Mg_8He$ and its Mg sublattice. (d)

Bader charge on ISQs in Mg$_8$He and its Mg sublattice. Please note there are 2 ISQIs and 7 ISQIIs in the primitive cell of Mg$_8$He. (e) The ELF for MgHe and its Mg sublattice, in which the large green and red balls represent Mg and He atoms, and the small blue and purple balls represent the position of ISQI and ISQII, respectively. Note that the ELF in He site shows an ambiguous value about 0.7. Nonetheless, we process it as ISQ. (f) The ELF for the Mg$_8$He and Mg sublattice of Mg$_8$He. The small blue and purple balls represent the ISQI and ISQII sites, respectively.

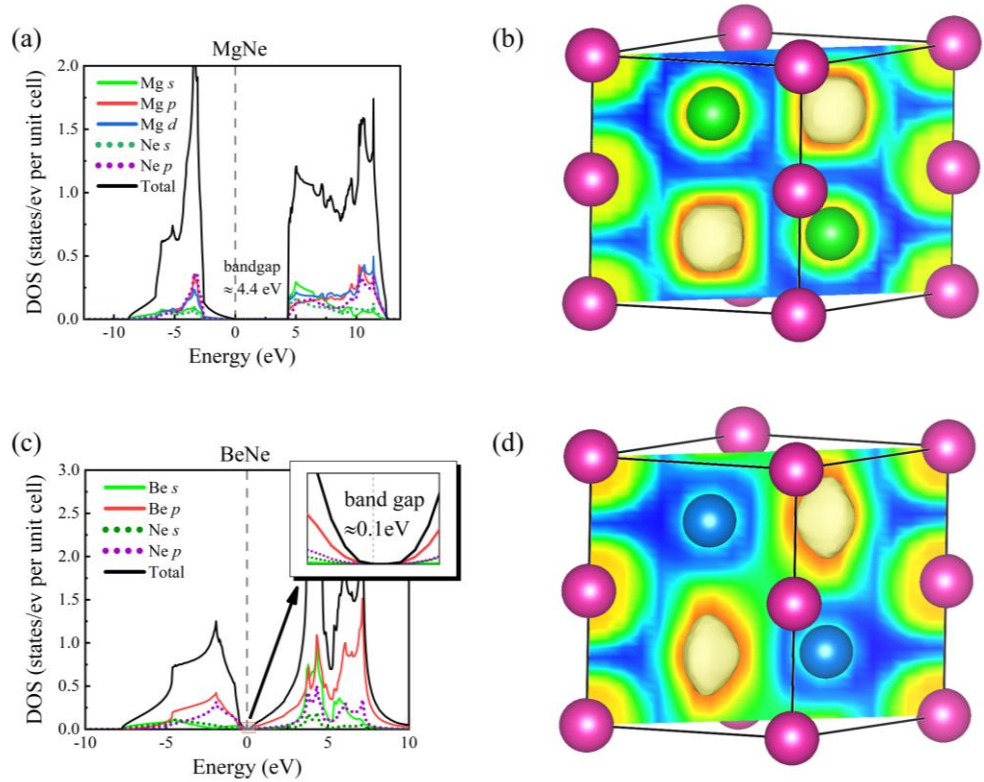

**Figure S7.** The crystal structure and ELF (b, d) and electronic DOS (a, c) of MgNe and BeNe at 500 GPa, respectively. The pink, green, and blue balls represent Ne, Mg and Be atoms, respectively. MgNe and BeNe are in the same structure as MgHe. The DOS of the MgNe and BeNe show an insulating and semiconducting behavior with a wide

bandgap of 4.4 eV and a narrow gap of 0.1 eV at the level of PBE, respectively. The opening of the bandgap in these two compounds are also originated from (i) the charge localization to ISQ, (ii) transfer of charge to NG atoms, and (iii) hybridization of the *s* and *p* orbitals of Ne with the corresponding metallic elements (*s*, *p*, *d* orbitals for Mg, and *s* and *p* orbitals for Be), which is similar to MgHe. The DOS of MgNe shows a much stronger hybridization than MgHe, and the bandgap is also wider. The ELF of MgNe and BeNe exhibit similar charge localization feature as MgHe, indicating their mixing feature of the ionic and covalent bonding. The slight overlapping of neighboring ISQs in BeNe (as shown in (d)) might be one of the reasons that lead to a narrow bandgap in this compound.

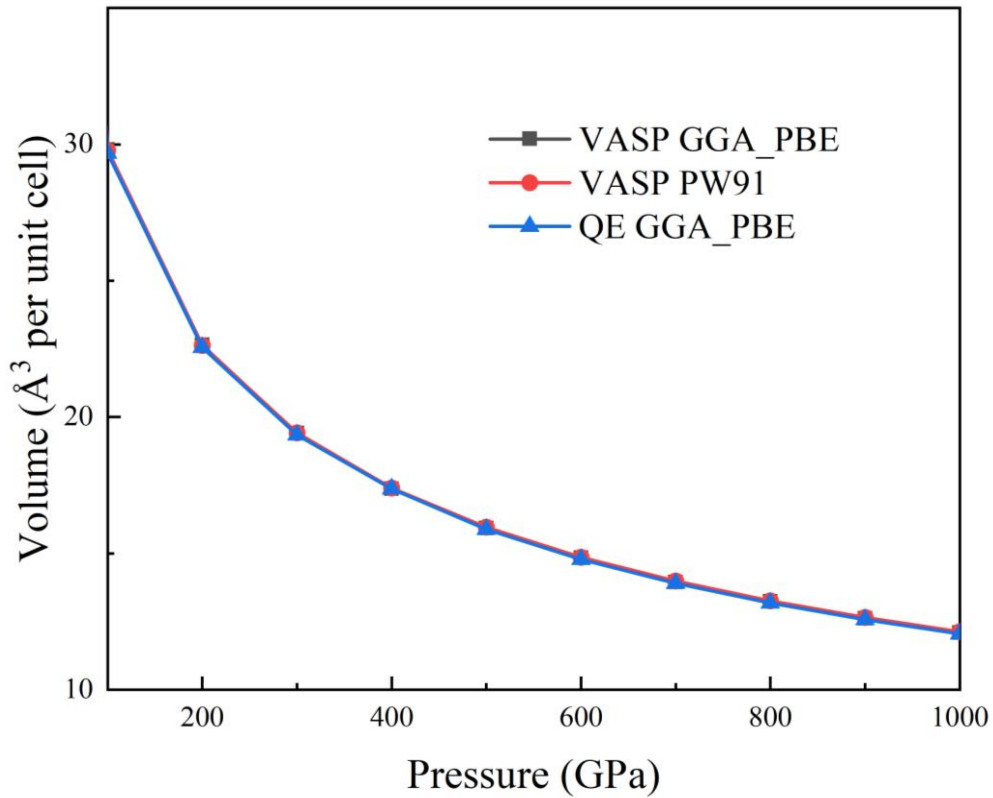

**Figure S8.** The equation of states (EOS) of MgHe calculated by using PBE and PW91

pseudopotential in VASP code and the PBE pseudopotential in Quantum Espresso (QE) code. In particular, the PBE potential used in QE has a small cut-off radius (0.5 Å), which is shorter than half of the interatomic distances, thus avoids core overlap (the minimal interatomic distance for each studied structure can be found in Supplementary Table S1). The comparison of these EOS suggests that the potential used in VASP code can give reasonable results within the pressure range of up to at least 1TPa.

**Table S1.** Structural parameters of the predicted and proposed stable compounds in He-Mg system at 1 TPa.

| Compounds | Space Group | Lattice parameter (Å, °) | Atomic positions | Minimal interatomic distance (Å) |
|---|---|---|---|---|
| MgHe | $P6_3/mmc$ | a=2.2795<br>b=2.2795<br>c=2.6818<br>α=90.0<br>β=90.0<br>γ=120.0 | Mg:(0.3333,0.6667,0.7500)<br>Mg:(0.6667,0.3333,0.2500)<br>He:(0.0000,0.0000,0.5000)<br>He:(0.0000,0.0000,0.0000) | Mg-Mg=1.880<br>Mg-He=1.478<br>He-He=2.281 |
| $Mg_6He$ | $R\bar{3}$ | a=6.1758<br>b=6.1758<br>c=3.0485<br>α=113.3<br>β=113.3<br>γ=113.3 | Mg:(0.0135,0.2458,0.4071)<br>Mg:(0.9864,0.7541,0.5928)<br>Mg:(0.4071,0.0135,0.2458)<br>Mg:(0.5928,0.9864,0.7541)<br>Mg:(0.2458,0.4071,0.0135)<br>Mg:(0.7541,0.5928,0.9864)<br>He:(0.5000,0.5000,0.5000) | Mg-Mg=1.797<br>Mg-He=1.487<br>He-He=3.708 |
| $Mg_8He$ | $R\bar{3}$ | a=4.1374<br>b=4.1374<br>c=8.4323<br>α=67.8<br>β=67.8<br>γ=67.8 | Mg:(0.1207,0.6963,0.3736)<br>Mg:(0.8792,0.3036,0.6263)<br>Mg:(0.3736,0.1207,0.6963)<br>Mg:(0.6263,0.8792,0.3036)<br>Mg:(0.6963,0.3736,0.1207)<br>Mg:(0.3036,0.6263,0.8792)<br>Mg:(0.8276,0.8276,0.8276)<br>Mg:(0.1723,0.1723,0.1723)<br>He:(0.5000,0.5000,0.5000) | Mg-Mg=1.687<br>Mg-He=1.475<br>He-He=3.689 |
| $Mg_{10}He$ | $P\bar{1}$ | a=4.1049 | Mg:(0.1220,0.1133,0.3863) | Mg-Mg=1.660 |
| | | b=3.6290 | Mg:(0.2821,0.7263,0.3137) | Mg-He=1.443 |
| | | c=3.6621 | Mg:(0.9776,0.3319,0.9410) | He-He=3.662 |
| | | α=67.1 | Mg:(0.7505,0.8127,0.6353) | |
| | | β=80.2 | Mg:(0.5264,0.1688,0.7006) | |
| | | γ=91.6 | Mg:(0.6708,0.9502,0.1459) | |
| | | | Mg:(0.8979,0.4694,0.4516) | |
| | | | Mg:(0.5124,0.3756,0.2041) | |
| | | | Mg:(0.1359,0.9065,0.8828) | |
| | | | Mg:(0.3663,0.5557,0.7732) | |
| | | | He:(0.8242,0.6410,0.0434) | |
| Mg$_{15}$He | $P\bar{1}$ | a=3.9885 | Mg:(0.2463,0.5594,0.0271) | Mg-Mg=1.662 |
| | | b=4.0315 | Mg:(0.2574,0.0648,0.2778) | Mg-He=1.445 |
| | | c=4.704 | Mg:(0.7605,0.8403,0.4062) | He-He=3.989 |
| | | α=89.6 | Mg:(0.9840,0.1835,0.9713) | |
| | | β=106.9 | Mg:(0.2352,0.0540,0.7765) | |
| | | γ=99.4 | Mg:(0.7321,0.2786,0.6481) | |
| | | | Mg:(0.0119,0.6576,0.6899) | |
| | | | Mg:(0.4706,0.8857,0.5932) | |
| | | | Mg:(0.4652,0.4430,0.8000) | |
| | | | Mg:(0.7760,0.7878,0.9104) | |
| | | | Mg:(0.0220,0.2331,0.4610) | |
| | | | Mg:(0.0274,0.6758,0.2543) | |
| | | | Mg:(0.5085,0.9354,0.0830) | |
| | | | Mg:(0.4807,0.4612,0.3643) | |
| | | | Mg:(0.7166,0.3310,0.1439) | |
| | | | He:(0.2463,0.5594,0.5271) | |
| Mg$_{18}$He | $P\bar{1}$ | a=5.2187 | Mg:(0.9881,0.3367,0.6520) | Mg-Mg=1.651 |
| | | b=3.6342 | Mg:(0.0640,0.7051,0.3167) | Mg-He=1.451 |
| | | c=5.0082 | Mg:(0.3549,0.7643,0.4460) | He-He=3.634 |
| | | α=71.9 | Mg:(0.6515,0.2188,0.6158) | |
| | | β=85.2 | Mg:(0.4581,0.1337,0.1100) | |
| | | γ=106.1 | Mg:(0.5177,0.8413,0.9297) | |
| | | | Mg:(0.6829,0.8960,0.4569) | |
| | | | Mg:(0.3183,0.1446,0.5766) | |
| | | | Mg:(0.8392,0.9428,0.9698) | |
| | | | Mg:(0.1672,0.0745,0.9806) | |
| | | | Mg:(0.4186,0.4771,0.2478) | |
| | | | Mg:(0.5341,0.5021,0.7746) | |
| | | | Mg:(0.7386,0.5802,0.2958) | |
| | | | Mg:(0.2038,0.6942,0.8500) | |
| | | | Mg:(0.7834,0.2586,0.1308) | |
| | | | Mg:(0.1035,0.3617,0.1789) | |
| | | | Mg:(0.8706,0.6200,0.8108) | |
| | | | Mg:(0.0044,0.9975,0.4970) | |
| | | | He:(0.2611,0.4194,0.7133) | |
| Mg$_{19}$He | $R\bar{3}m$ | a=5.4017 | Mg:(0.5000,0.0000,0.5000) | Mg-Mg=1.632 |
| | | b=5.4017 | Mg:(0.9128,0.9128,0.3530) | Mg-He=1.460 |
| | | c=5.4017 | Mg:(0.3530,0.9128,0.9128) | He-He=5.401 |
| | | α=51.7 | Mg:(0.7867,0.7867,0.7867) | |
| | | β=51.7 | Mg:(0.2978,0.8169,0.2978) | |
| | | γ=51.7 | Mg:(0.7021,0.7022,0.1830) | |
| | | | Mg:(0.1830,0.7022,0.7022) | |
| | | | Mg:(0.5967,0.5967,0.5967) | |
| | | | Mg:(0.9128,0.3530,0.9128) | |
| | | | Mg:(0.0871,0.6469,0.0871) | |
| | | | Mg:(0.5000,0.5000,0.0000) | |
| | | | Mg:(0.0000,0.5000,0.5000) | |
| | | | Mg:(0.4032,0.4032,0.4032) | |
| | | | Mg:(0.8169,0.2977,0.2977) | |
| | | | Mg:(0.2978,0.2977,0.8169) | |
| | | | Mg:(0.7021,0.1830,0.7021) | |
| | | | Mg:(0.2132,0.2132,0.2132) | |
| | | | Mg:(0.6469,0.0871,0.0871) | |
| | | | Mg:(0.0871,0.0871,0.6469) | |
| | | | He:(0.0000,0.0000,0.0000) | |
| Mg$_{27}$He | $R\bar{3}m$ | a=7.1487 | Mg:(0.9332,0.9332,0.4245) | Mg-Mg=1.665 |
| | | b=7.1487 | Mg:(0.8525,0.8525,0.8525) | Mg-He=1.461 |
| | | c=7.1487 | Mg:(0.5000,0.0000,0.5000) | He-He=7.149 |
| | | α=38.4 | Mg:(0.8111,0.8111,0.2506) | |
| | | β=38.4 | Mg:(0.4245,0.9332,0.9332) | |
| | | γ=38.4 | Mg:(0.7155,0.7155,0.7155) | |
| | | | Mg:(0.0000,0.5000,0.5000) | |
| | | | Mg:(0.9332,0.4245,0.9332) | |
| | | | Mg:(0.3690,0.8420,0.3690) | |
| | | | Mg:(0.6309,0.6309,0.1579) | |
| | | | Mg:(0.2506,0.8111,0.8111) | |
| | | | Mg:(0.5818,0.5818,0.5818) | |
| | | | Mg:(0.8420,0.3690,0.3690) | |
| | | | Mg:(0.5000,0.5000,0.0000) | |
| | | | Mg:(0.8111,0.2506,0.8111) | |
| | | | Mg:(0.1888,0.7493,0.1888) | |
| | | | Mg:(0.1579,0.6309,0.6309) | |
| | | | Mg:(0.4181,0.4181,0.4181) | |
| | | | Mg:(0.7493,0.1888,0.1888) | |
| | | | Mg:(0.3690,0.3690,0.8420) | |
| | | | Mg:(0.6309,0.1579,0.6309) | |
| | | | Mg:(0.0667,0.5754,0.0667) | |
| | | | Mg:(0.2844,0.2844,0.2844) | |

| |
|---|
| Mg:(0.5754,0.0667,0.0667) |
| Mg:(0.1888,0.1888,0.7493) |
| Mg:(0.1474,0.1474,0.1474) |
| Mg:(0.0667,0.0667,0.5754) |
| He:(0.0000,0.0000,0.0000) |